# Mathematical basis for polySAT implication operator

Charles Sauerbier
cs@orst.edu

ABSTRACT — The mathematical basis motivating the "implication operator" of the polySAT algorithm and its function is examined. Such is not undertaken with onerous rigor of symbolic mathematics; a more intuitive visual appeal being employed to present some of the mathematical premises underlying function of the implication operator.

KEY TERMS — Algorithms, complexity, computation theory, satisfiability, group theory, field theory, set theory.

## I. INTRODUCTION

The problem of determining a solution to Boolean expressions has been a noteworthy puzzle for more than a few years. Stephen Cook's formal proposition of the complexity of such as the satisfiability problem, [1], has added to this puzzle the question of whether there even exists means to decide the existence of a solution to any arbitrary instance of simple Boolean expressions having form without hindrance of quantifiers. [2] proposed an algorithm (polySAT) in solution of Cook's challenge that determines both existence and instance of a solution for any arbitrary Boolean expression of Cook's form – the satisfiability problem. This paper examines the "implication operator" of the polySAT algorithm, in effort to give greater clarity to its motivation and function.

The mathematical concepts that form the basis by which the implication operator of polySAT is motivated are presented here without onerous rigor. This is done in part because such is seen as not of necessity to the purpose of this paper. It is also hoped this paper will be more approachable to those for whom the major themes are not fields of expertise. The author makes no pretense for such expertise as well.

## II. CONSTRUCTIVE DEVELOPMENT OF IMPLICATION OPERATOR

### A. Convention and Intuition

It is human nature to trust convention. Convention arguing against a polynomial time solution of Cook's problem, existence of such solution seems counter-intuitive. The motivation by which the algorithm derives is likewise perhaps rather counter-intuitive to most observers. For the general topic of complexity and as reference to Cook's satisfiability problem [3] may likely be a more familiar and handier reference. Our purpose here is not to cover material that has become common knowledge in over 30 years, except where some subtle or obscure fact or detail is required, we shall not be littering the text with references to the ordinary.

An instance of SAT being a set of variables, represented by U, taking values over the set {F, T}, and; an expression, represented by E, in U of form $c_1 \wedge c_2 \wedge \ldots \wedge c_n$, a conjunction of clauses $c_i$, where C represents the set of all $c_i$ in E, and; all $c_i$ in E are of form $u_1 \vee u_2 \vee \ldots \vee u_n$, a disjunctive clause in U. The instance of SAT being represented by E = (U, C).

It is easily seen that the set {F, T} is equivalent to $Z_2$ to within a homomorphism. It follows then that for each variable $u_i \in U$, for any instance of SAT, the values $u_i$ is allowed to assume exist as well as an element within a vector space of one dimension over $Z_2$. To facilitate discussion, let the vector space be visually represented as depicted in the illustration given in Fig. 1a (below).

*B. Vector Spaces, Partitions, Groups, and Fields*

A one dimensional vector space over $Z_2$ may be used to represent the set of all possible assignment of values to any variable $u_i$, but does not contain any information indicating which values $u_i$ is not allowed to assume and is allowed to assume. For this we need to extend the representation by some means if we desire to represent such information within the vector space model. Clearly by definition a point within the vector space cannot be both disallowed and allowed by simple consequence of definition of these terms. The sets *disallowed* and *allowed* must then form a partition – set of disjoint sets – of the set of elements within the vector space. Let the set of colors {RED, GREEN} be used to visually represent whether an element in the vector space is disallowed (RED) or allowed (GREEN) as an assignment to $u_i$. It follows then that the vector space has 4 possible partitions, which are illustrated in Fig. 1b.

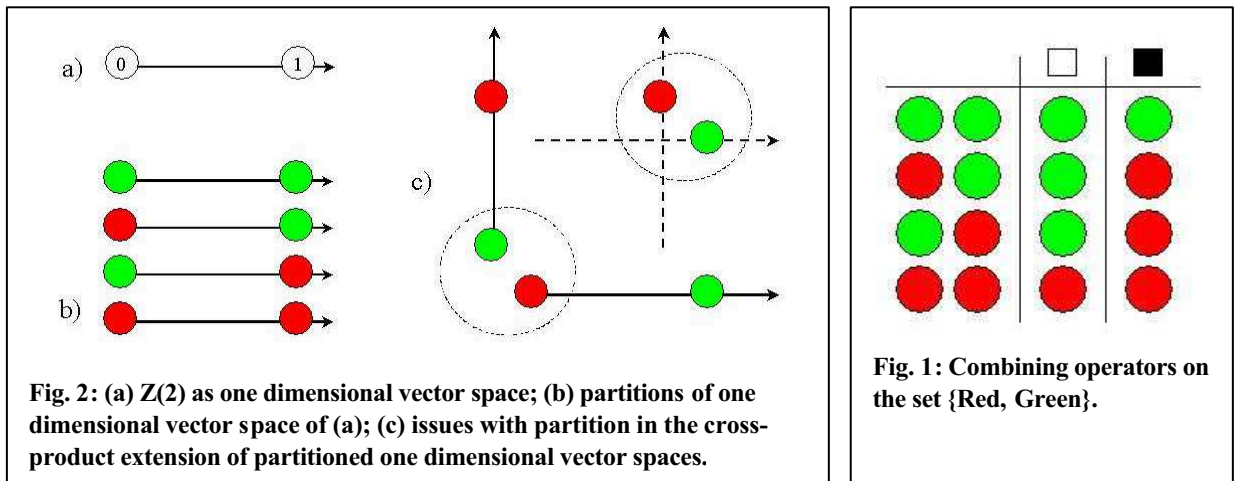

Fig. 2: (a) Z(2) as one dimensional vector space; (b) partitions of one dimensional vector space of (a); (c) issues with partition in the cross-product extension of partitioned one dimensional vector spaces.

Fig. 1: Combining operators on the set {Red, Green}.

Given suitable vector spaces larger vector spaces may be constructed by "combining" the underlying fields of two or more vectors spaces. A vector space also may be extended by combining the underlying field with itself. One way of combining suitable vector spaces is to form the cross product of the elements of the underlying fields of the vector spaces ("ordinary" field extension). Without restriction on how a vector space is extended, in extending it we are confronted by the question of how to resolve the partition between disallowed and allowed of elements of the new vector space thus constructed (Fig. 1c).

The resolution of the issue of the partition ultimately depends on objective outcome. In the context of motivation for polySAT the resolution of the partition issue is considered for the cross-product of vector spaces using two operators symbolically represented by *white square* and *black square* symbols in Fig. 2, where the associated rules of combination are also defined. Let **WS** be the textual representation of the white square operator, for convenience. Let **BS** be the textual representation of the black square operator, for convenience.

It may not be immediately obvious, but the operators, **WS** and **BS**, both have the properties of closure (see Fig. 2), identity (RED for both), associative, and commutative; only **BS** has an inverse (RED). The operators, **WS** and **BS**, are also distributive over one another. By definition then ({RED, GREEN}, **WS**) is a monoid, and ({RED, GREEN}, **BS**) is an abelian or commutative group. ({RED, GREEN}, **WS**, **BS**) is therefore a commutative ring with identity by definition. Since it is a commutative ring with an identity it is a field by definition. It follows that the result of this is to have constructively defined a field over a partition of a field.

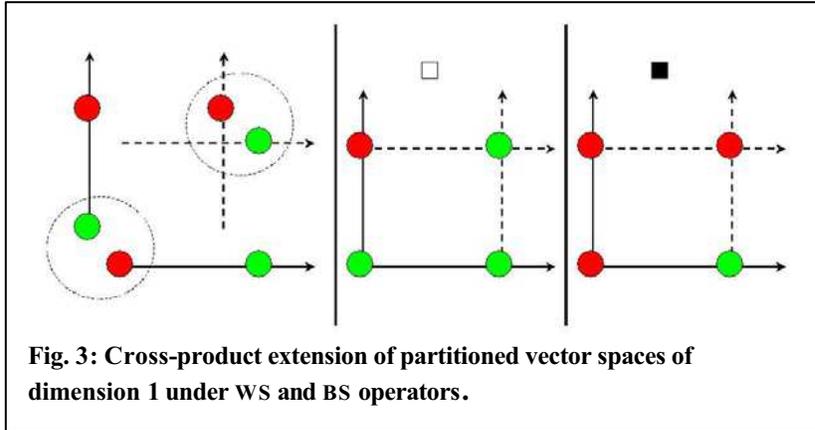

Fig. 3: Cross-product extension of partitioned vector spaces of dimension 1 under **WS** and **BS** operators.

Applying either operator to the cross-product of two vector spaces of dimension 1, subject to a partition, it is now possible to construct a vector space of dimension 2 for which the partition is an image, under either the **WS** or the **BS** operator, of the partition of each of the two vector spaces in the resultant vector space constructed by the cross-product of the underlying fields, as illustrated in Fig. 3. Applying the operators to cross-product of a vector space of dimension 1 with a vector space of dimension 2, a vector space of dimension 3 can be constructed with an attendant partition, as illustrated in Fig. 4. As an issue of semantics, let **WS** define the property "GREEN preserving", and let **BS** define the property "RED preserving"; the motivation of which is presumed visually evident.

*C. Combinatorial Operations on Partitioned Vector Spaces*

Taking the cross-product is one means to effect a combinatorial operation on vector spaces. It is not our only means of doing so, however. Consider applying the operators **WS** and **BS** to vector spaces of dimension 3 on basis of the intersection of the dimensions of each respective vector space. As a matter of first-impression the purpose and benefit of such may not be obvious, or even intuitive, until one considers that every vector space, $\vec{v}$, over $Z_2$ contains within it a set of vector subspaces, defined by the power set, $\wp(v)$, of the scalar coordinate set, $v$, of $\vec{v}$. It follows then that by applying the operators **WS** and **BS** to element vector subspaces of any subset of $\wp(v)$, for which a partition is defined for each element and the union of scalar coordinate set of all vector subspaces in the set is equivalent to the scalar coordinate set, $v$, of $\vec{v}$, it is possible to define a partition of $\vec{v}$, as well as to construct $\vec{v}$ in the process. It is further posited that any desired partition of $\vec{v}$ can thus be defined. The later proposition, as well as that of construction of $\vec{v}$ in the prior, not being of significance to the issues at point for this paper, proof of it is not here included.

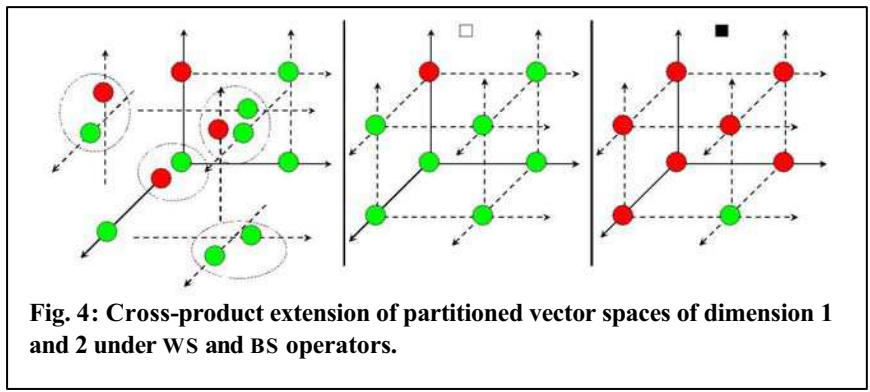

Fig. 4: Cross-product extension of partitioned vector spaces of dimension 1 and 2 under **WS** and **BS** operators.

If it is desired to apply the operators **WS** and **BS** as combinatorial operations to partitioned vector spaces there is a need to define how such is to be accomplished. It is observably evident that two vector spaces of dimension 3, differing by at least one scalar coordinate, can intersect only in one of the faces of the defined brick or one of the coordinate axis. (The "virtual cluster" relation observable between elements of the clausal partition, thus clauses in [2], for any instance of 3SAT, is a corresponding representation of such intersection.) The objective intent of application of the operators **WS** and **BS** as combinatorial operations not being strictly the resolving of the partition of elements in two vector spaces into that of a single vector space, let a white circle (**WC**) and black circle (**BC**), respectively, represent the desired application of the operators **WS** and **BS** in the context of combinatorial operations on vector spaces. Only **BC** being of actual concern to constructive understanding of the implication operator of polySAT, we will leave **WC** for another day to consider; defining here how to effect **BC** on two partitioned vector spaces and limiting our definition to vector spaces of dimension 3.

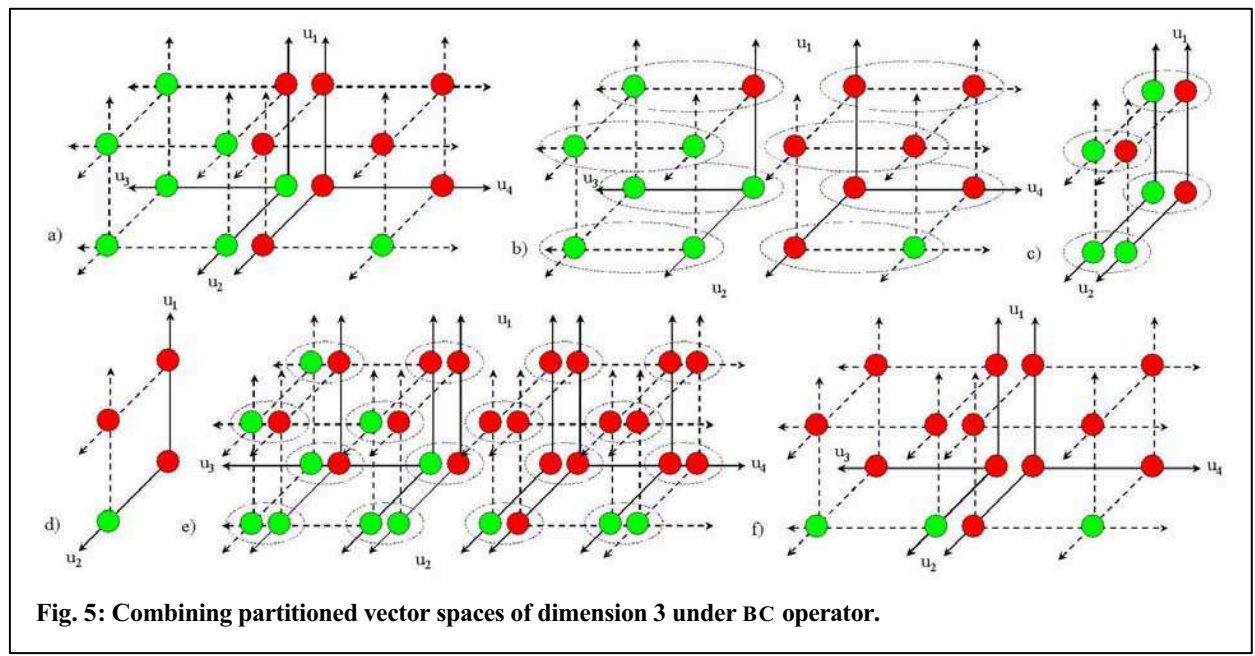

**Fig. 5: Combining partitioned vector spaces of dimension 3 under BC operator.**

Figure 5 visually defines the process by which **BC** is accomplished for partitioned vector spaces of dimension 3. Fig. 5a depicts the two vector spaces to which **BC** is to be applied. First the vector spaces are mapped to a vector space defined by the intersection of their scalar coordinates. The operation is being depicted for two vector spaces intersecting in two of their respective scalar coordinates, thus each is mapped to a vector space of dimension two. The mapping is defined by first applying **WS**, the GREEN preserving operation, so as to combine elements in a vector space along vectors orthogonal to axis of scalar coordinates participating in **BC**. The dashed circles in Fig. 5b indicate the pairs to which **WS** will be applied. Where the vector spaces intersect only in a single scalar coordinate, the **WS** operator would again be applied to map the result to a vector space of one dimension. Next the **BS**, RED preserving, operator is applied to the resulting pair of lower dimensional vector spaces, Fig. 5c, to produce a single partition for the vector space, Fig. 5d. The resulting single partition is then applied to the two vector space operands by performing the **BS** operation on it in combination with the elements in each vector space, Fig. 5e, by indexing over the scalar coordinate(s) not in the intersection of such between the vector spaces. Fig. 5f gives the results of applying **BC** to the vector spaces of Fig. 5a.

## D. Partitioned Vector Spaces and Clausal Partitions

The clausal partition, D, defined as part of the polySAT algorithm, is equivalent to the set of partitioned vector spaces of the corresponding elements of $\wp(v)$, where $v$ is equivalent to U to within a homomorphism and each takes values from some set, respectively, that are also equivalent to within a homomorphism. Each element of the clausal partition is defined in terms of a 3-tuple of variables. The union of the defining 3-tuples of variables for all elements in D is U, therefore the scalar coordinates of the corresponding vector spaces covers $v$ where $v$ and U are of equal cardinality. It is clearly apparent that both {F, T} and {RED, GREEN} are equivalent to $Z_2$ to within a homomorphism, thus to one another to within a homomorphism. Finally, as the complement, $\bar{c}$, of each clause, c, defined in SAT is disallowed as satisfying assignment of the corresponding element, d, of the clausal partition. In example: Consider the clause $\bar{u_1} \vee u_2 \vee \bar{u_3}$, viewed as it binary representation (F, T, F), its compliment, (T, F, T), cannot satisfy any clausal partition in which the clause exists, as the clause will not satisfy with such assignment; thus the compliment is disallowed as an assignment. The equivalence is thus made evident.

## E. Partitioned Vector Spaces, Clausal Partitions, and Boolean expressions using OR and AND

Perhaps not immediately obvious is the equivalence of the result of **BC** to the conjunction of clauses in 3SAT. The irrefutable proof of this being to laboriously evaluate all possible results of **BC** together with the corresponding solutions of conjunctive expressions over two disjunctive clauses on 3 variables where such have two variables in common. Like wise for the case where the disjunctive clauses have a single variable in common.

Exhaustive proof being far too lengthy for inclusion here, it is omitted. Instead, observe that the definition of the operator **WS** is equivalent to $\vee$, and **BS** to $\wedge$, by respective definitions. It is then immediately obvious on comparative examination of the truth tables of two disjunctive clauses with the resultant truth table of the conjunction that the net effect is the reduction of the allowable assignments in each where either of the conjunctive clauses does not admit the assignment to the variables in common between the two clauses. The net effect of the mapping step performed with the **WS** operator is to produce a partition over the common variables reflecting the disallowed and allowed assignments by means of partition. The net effect of the subsequent imposition of that resultant partition into each vector space under the **BS** operator is to mark as disallowed (RED) all instances in each, respectively, having disallowable assignment to the scalar coordinates, thus variables, in common between the two vector spaces (clauses).

The foregoing is observably more relevant were one considers the equivalence of the **BC** operator with the conjunction of disjunctive clauses in the elements of the clausal partition of 3SAT, as the later is defined in [2] for polySAT, and the operation of the implication operator defined for polySAT. Fig. 6a gives the binary representation of the consequence of having applied the **BC** operation to the vectors of Fig. 5 (see Fig. 5f). Fig. 6b gives the binary representation of equivalent partition of a single vector space having four scalar dimensions of two vector spaces to which **BC** was applied. Fig. 6c gives in binary representation the clausal partitions corresponding to the vector spaces given in Fig. 5a, under the assumption that such represents initial state of clausal partitions in polySAT algorithm. It is easily verified that the conjunction of the set of clauses given in Fig. 6c is satisfied only by those assignments marked as allowed (GREEN) in either of Fig. 6a or Fig. 6b.

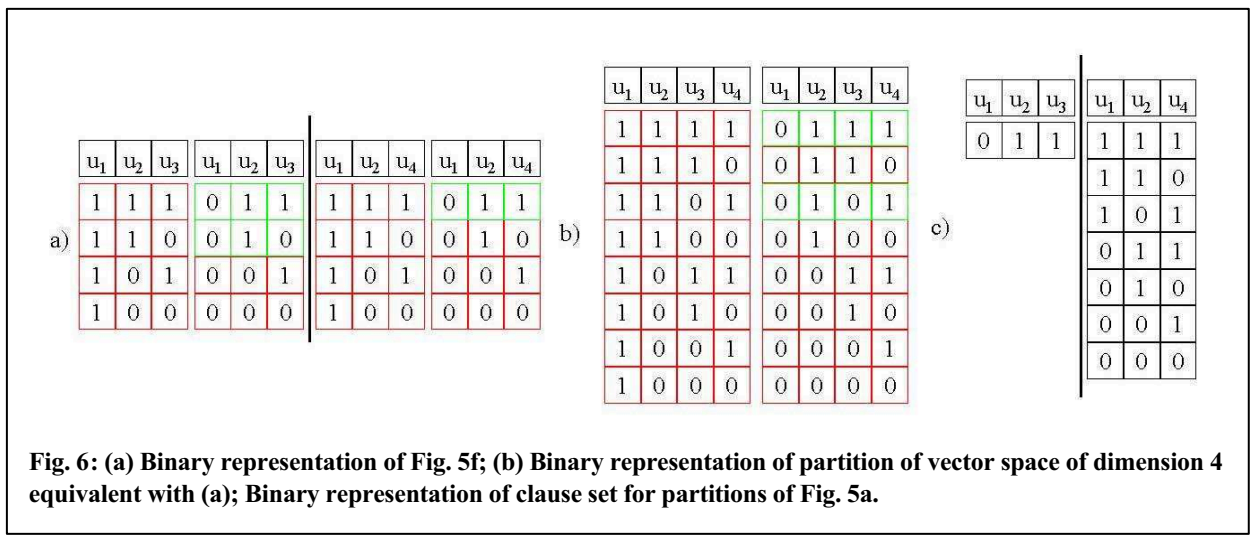

**Fig. 6: (a) Binary representation of Fig. 5f; (b) Binary representation of partition of vector space of dimension 4 equivalent with (a); Binary representation of clause set for partitions of Fig. 5a.**

*F. Unidirectional Implication Operator of polySAT*

The implication operator as defined and used in polySAT is unidirectional; that is to say in context of the **BC** operation on partitioned vector spaces, the implication operator imposes the partition of one vector space onto another. This is as if one skipped the **BS** operation depicted in Fig. 5c and simply performed the subsequent **BS** operation on one vector space using the lower dimension partition map from the other vector space. The legitimacy of this is evident where one performs this "unidirectional-**BC**" operation again with the role of each partitioned vector space reversed. As with the unidirectional implication operator vs bi-directional implication operator, the unidirectional-**BC** needs to be applied until the system settles to steady state for the result to be assured equal to the result of "bi-directional" **BC** as defined above. The unidirectional performance of **BC** in manner of the unidirectional implication operator in polySAT is illustrated in Fig. 7 (below).

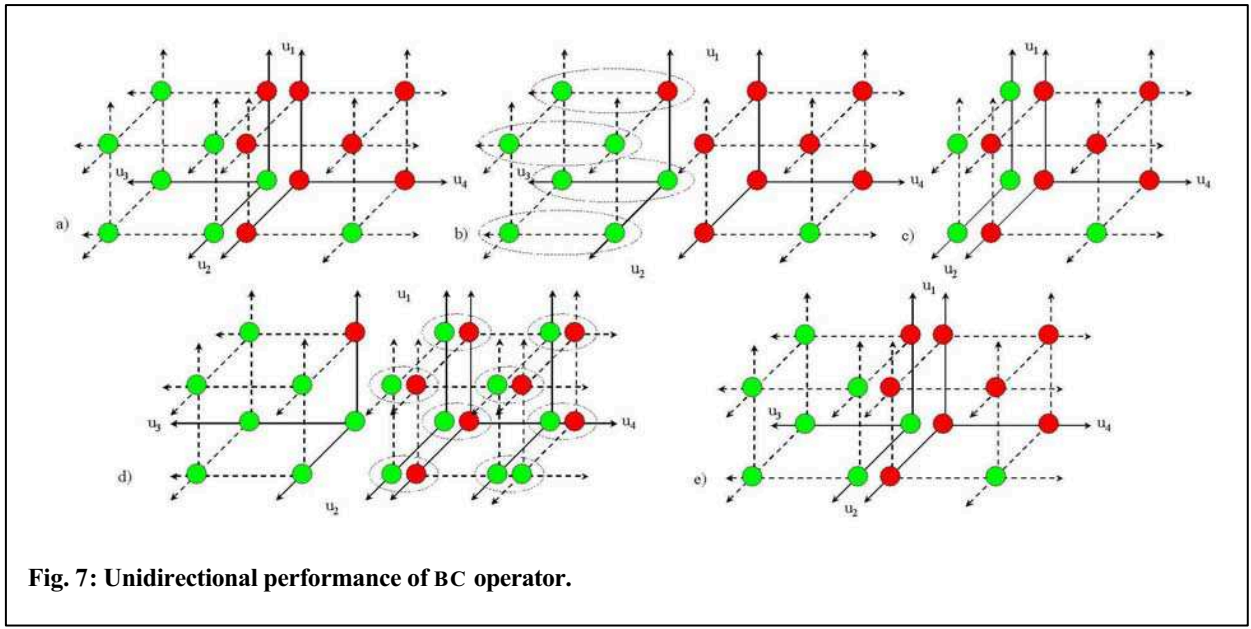

**Fig. 7: Unidirectional performance of BC operator.**

## III. RELATION TO WORK OF OTHERS

A brief discussion of findings respective field extensions of vectors spaces under constraint of coupling between fields is presented in this paper. The body of work in group theory, field theory, and abstract algebra, generally, is extensive, and; the author makes no claim of expertise in these fields, nor complete knowledge of the body of literature of such. With such disclaimer, the author is unaware of works by others as such relates to extension of fields under constraints as discussed in this paper. References to such works as others may have related to more than the ordinary field extension would be most welcomed.

## IV. CONCLUSION

The mathematical basis motivating the implication operator defined and used by polySAT was presented absent onerous rigor. The purpose of the paper being not to present substantive new details of the polySAT algorithm as much as to illuminate details previously given in [2], highlighting the mathematical motivations of the polySAT implication operator as prelude to broader examination of the dynamic effect of the implication operator in the context of polySAT. The mathematical properties of the presented operator BC is likewise deferred for later examination as an ongoing issue of research.